# How they (should have) built the pyramids

Dr. J. West*, G. Gallagher*, K. Waters**
*Department of Chemistry and Physics, Indiana State University, Terre Haute, IN 47809
**Department of Physics, Michigan Technological University, Houghton MI 49931
(conducted while an undergrad at ISU)

Abstract

A novel method is proposed for moving large (pyramid construction size) stone blocks. The method is inspired by a well known introductory physics homework problem, and is implemented by tying 12 identical rods of appropriately chosen radius to the faces of the block. The rods form the corners and new faces that transform the square prism into a dodecagon which can then be moved more easily by rolling than by dragging. Experimental results are presented and compared to independent work by another group which utilized wooden attachments providing a cylindrical shape. It is found that a small scale stone block converted to dodecagons can be moved across level open ground with a dynamic coefficient of friction of the order of 0.2. For full scale pyramid blocks, the wooden "rods" would need to be posts of order 30 cm in diameter, similar in size to those used as masts on ships in the Nile.

## I. INTRODUCTION

Some of the pyramids in Egypt, architectural wonders, were built more than 4000 years ago. As a basis for the scale of the effort involved, it is noted that there are about 2.3 million blocks that make up the pyramid of Khufu.[1,2] Those blocks were quarried locally from limestone (average density 2.2 kg/m$^3$) with average dimensions 1.27m x 1.27m x 0.71m and imposing mass of 2500kg. The blocks, once quarried, were moved approximately 2 km to the construction site, and then into position, some to the top of the pyramid at an altitude of 138 m.[1-3] The method of construction of these impressive monuments has been a subject of intensive investigation, and speculation by scholars from a wide range of disciplines, and armchair engineers alike.

There have been a number of documentaries produced on the pyramids in general, and some have included large scale tests of possible construction methods. Of particular interest is a documentary episode produced by NOVA in which Mark Lehner and Roger Hopkins actually lead a team that built a small pyramid of approximately 10 m in height.[4] They tested moving the blocks by dragging them on sleds, or on sleds with some lubrication, over prepared roadways. In the experiments of Fall et. al.[5] the results indicate that the percent water content of the sand (about 5% was optimal) decreasing the "dynamic friction $\mu_d$, by as much as 40%. The type of sand used also had an effect, but the lowest value reported was $\mu_d = 0.30$. An impressive reduction, but still a relatively high value. More importantly, not all of the pyramids were built on the banks of a major body of water, so that the transport of the lubricant in the form of the water might become a project as difficult as the moving of the blocks.

As an alternative to dragging large blocks, one can consider rolling the blocks. Rolling a prism of 4 sides is not efficient, but adding wooden rods to the surface can effectively increase the number of sides. The crew can then pull on a rope wrapped around and passing over the top of the block. In this configuration, static friction acts in the direction of the desired motion, rather than opposing the motion. In effect the block and rope combination becomes a 2:1 pulley, though the pulley was not yet formally "known" to the Egyptians at that time.[1,3] The energy loss due to friction when rolling small scale blocks across open ground in this configuration is equivalent to a coefficient of $\mu_d = 0.3 \pm 0.10$, while the force required to maintain the motion of the block is only (mg$\mu_d$)/2, or about 15% of the weight of the block. The rods can be reused many times, and there is no need to to transport large quantities of water for lubrication.

Independent work appearing in the book *Engineering the Pyramids* by D. Parry[6] describes full scale experiments involving a 2.5 tonne stone with dimensions 0.8 m x 0.8 m x 1.6 m.[6] The stone was moved 80 meters on level ground (using 160 m of rope), and up a 1 in 4 slope, steeper than the 1:12 or 1:8 grade ramps that are in evidence at some of the pyramid sites.[1,3] The experiments by Parry and coworkers utilized wooden quarter circle "cradle runners" or "rockers" mentioned in other texts, but not in this context.[3] The rockers are attached, one to each face of the block, by simply wrapping rope around them, turning the square prism block into a cylinder. The experiments by Parry indicate "that a block could be rolled at a fast walking pace by two or three men pushing it from behind." Much faster than the 4 meter / minute rate assumed by Smith, and a much smaller crew with much less effort than required in "This Old Pyramid." The same block could be moved short distances up the 1:4 slope by 10 men, and was moved up the full 15 m ramp by 14, although a crew of 20 was more routinely used. Tension gauges were not used, but the number of men required to move the block would suggest the work required is consistent with a coefficient of friction of $\mu_d = 0.2$ on a prepared roadway. Within the book by Perry, earlier tests by "an American engineer" named John Bush and carried out in Boston in 1980, using similar quarter circles are mentioned. Online sources place this work by Bush in 1977,[7] or 1978,[8] but no specifics about the experiments are supplied, and the results are not "published" as far as the authors can tell.

An advantage of rods over the rockers, and even more so over dual runner sleds is a reduction in ground pressure which would greatly reduce the maintenance required for the road surfaces, or possibly do away with the need for prepared surfaces entirely. The diameter of the rods required are consistent with the diameter of the masts used on ships in Egypt at that time,[9,10] the same ships that would be employed in moving granite blocks along the Nile.

An interesting rolling variant can be accomplished by using catenary shaped wedges of wood to change the shape of the road, instead of the shape of the blocks. Again, from a book written by a civil engineer, comes the idea to change the shape of the ground (locally) rather than the shape of the blocks, by way of wooden "humps" with catenary cross sections.[11] Modifying a roadway in such a manner is an idea that has been known for some time. A particularly nice example using a square wheeled tricycle is found on and SPS chapter website.[12] The method clearly works for blocks as well,[11] but requires the fabrication of many of the specially shaped road segments, and it is not clear that the road segments would be stable on inclines or in rainy conditions.

The remainder of the paper is organized as follows: Section II presents a short review of the effect by which friction is used to help move the blocks when rolling them, Section III describes the experimental methods, specifying the radius needed for the rods, the method of attaching the rods, and the experimental results. The conclusions are presented in Section IV.

**II. THEORY**

In dragging a block, it is clear that the work performed by the crew is lost as heating due to sliding friction. In rolling the block with cylinders attached, energy is also lost, but now due to the inelastic collisions of each face of the newly formed polygon with the ground. Most of the work performed in raising the center of mass from the face to a vertex is lost as it "falls" from the vertex onto the next face in an inelastic collision with the ground. However, not all of the mechanical energy is lost. The block will tend to roll onto the next vertex up a slight but noticeable amount (roughly estimated by eye to be an angle of 10 degrees) to the next vertex. This is, effectively, a form of rolling friction, and the primary source of an effective dynamic coefficient of friction $\mu_d$ for the polygon. The more sides the polygon has, the less energy is lost to collisions, up to a limit where actual "rolling friction" would begin to dominate. The crew works at roughly a constant rate, applying roughly a constant "tension of rolling" (TR) in order to counteract those energy losses. A complicating issue is that a larger "vertex tension" (TV) is required to get the polygon to the first vertex to initiate rolling. An elementary analysis based on forces and torques shows that the ratio TV/TR decreases for a polygon of N sides with increasing N as $\tan(\pi/N)$, approaching 1 as expected in the limit of a cylinder (large N). For the conditions of the experiments described here, TV/TR is of the order of 1.5 for some designs, as discussed below.

Rolling a cylinder with a rope wrapped over the top of the object is a well-known example/homework problem commonly addressed in first semester freshmen level physics courses.[1,14] It is used because of the "surprise" result that the force of friction is in the **same direction** as the applied tension, so that the acceleration a, of the cylinder is given by

$$a = \frac{4}{3}\frac{T}{m} < 4\mu_s g \tag{1}$$

where T is the tension in the rope and m is the mass of the cylinder. The magnitude of the acceleration is **greater than** T/m, while limited by possible slipping via $\mu_s$, the **static** coefficient of friction. This result neglects rolling friction (and face impacts for the dodecagon), but demonstrates the inherent advantage to passing the rope over the top of the block. The definition for $\mu_d$ used here to compare with previous work by others is

$$\mu_d = \frac{2T}{mg}. \tag{2}$$

which is the ratio of the work done in moving the block (and the attached rods) a given distance x, divided by the amount of work that would be required to lift the block a distance x/2 (recall that the crew moves twice as far as the block). This definition gives $\mu_d$ in terms of the mass of the block, and the average tension, each of which is easily determined by the experiments.

### III. EXPERIMENTAL METHODS AND RESULTS

Experiments were conducted using a concrete block (40 cm x 19.6 cm x 20 cm, mass = 29.6 kg, density = 2040 kg/m$^3$), and tested for consistency by with a second block of almost identical dimensions. A set of three identical wooden "dowel rods" were attached to each face of the block parallel to the axis of symmetry using rope, transforming the square prism into a 12-sided polygon (a dodecagon). The "dowel rods" were purchased at the local hardware store for less than $10 total cost. The required rod diameter d, is the difference between D, the distance from the center of the block to a vertex, and b, the distance from the center of the block to the center of the appropriate face of the dodecagon of the same center to vertex distance D. By an extremely useful mathematical coincidence, three rods of that diameter give a surface of width 2d, which is exactly the length of a side on a dodecagon with the same center to vertex distance as the square. Using the relationship between the vertex distance D, and the distance to a face on a dodecagon, it is straightforward to show that

$$d = D\cos(\pi/12) - 0.707D = 0.366b. \tag{3}$$

Each set of three rods becomes a new "face" of the dodecagon, while the 8 remaining "faces" are formed by the line connecting the square vertex to the edge of the attached rod faces. Each face is of length 2d = 0.73b (see Fig. 1). For the blocks used, the desired rod diameter is d = 3.66 cm, but the diameter of the rods chosen were d = 3.80 cm, the closest standard size dowel rod available (1.25, 1.5, and 2.0 inch).

Each set of three rods were lashed together side-by-side into a single "mat," and placed on the appropriate prism face. At each end of the block, the mats on opposite sides of the block were then attached to each other with a loose lashing. A short thin rod was inserted into the lashing and twisted, dramatically increasing the tension, and holding the four mats in place. An image of a block with the rods attached is shown in Fig. 2. The apex of the outside rod on each face forms 8 of the 12 dodecagon vertices. Three identical rods, all lashed together, was found to be a very stable configuration.

A heavy gauge string was wrapped part way around the block and rod body such that the person pulled on the rope over the top of the block and rod body. On runs for which data was recorded, the rope was laid out along the desired path, and the block and rod body was placed on top of the rope very near to one the end of the rope. That short end of the rope was then used to roll the block in the direction of the other end of the rope. The block was moved a total distance of almost one half of the length of the rope. Tension and position data were obtained as functions of time utilizing a Vernier Dual Range Force Sensor (order code DFS-BYA), and ULI Motion Detector (order code MD-ULI-O) connected via a DIN converter (CBL-RJ11) both connected to a Vernier LabQuest (order code LabQ) running Logger Pro (2.1) software. In each case, the block is moving at a constant speed of approximately 1.0 m/s as determined by video analysis of the experiment. The most realistic tests were conducted on the infield section of a regularly maintained practice softball field on the campus of Indiana State University (ISU). Some experiments were also conducted in the grass area of the outfield of this same softball field, and in a rough gravel of the parking lot. The field conditions were dry, other than the small amount of water used for regular maintenance, as the experiments were conducted during drought conditions. Experiments were also carried out indoors on a sheet of particle board approximately 0.80 m wide and 3.0 m long. The results obtained are shown in Fig. 3.

The regular rolling motion obtained indicates that the effective value of $\mu_d = 0.30$ ± 0.05 for steady rolling motion. A very fit human using the major muscle groups can perform long term physical work at a rate of approximately 75 W.[15] In order to maintain a constant block speed of 0.5 m/s a block of mass 2500 kg would require a work crew of roughly 50 individuals. In order to maintain the motion, each rope can be tied in a loop, but there would need to be a regular rotation of crew members from the front of the rope to the back of the rope due to the 2:1 crew to block distance disparity.

The indoor trials showed the greatest value of TV/TR = 2.0. As such, initial experiments were conducted in which two more rods of diameters less than d were attached to each block face (8 additional rods), with the smaller two rods fitting into the "slots" between the three identical rods already in place on each face. This makes the rod a 20-sided polygon, but not a regular polygon. The addition of the smaller rods reduced the ratio to TV/TR = 1.50 TR. A theoretical method of predicting optimal radii for these smaller rods is currently lacking and only these preliminary results are reported here. The additional rods did not produce a noticeable difference in the result obtained on the outdoor surfaces, nor did they produce a significant decrease in the energy lost once rolling was initiated.

## IV. CONCLUSIONS

A novel method for moving large stone blocks has been presented. The method involves the use of wooden rods applied externally to the stone block to facilitate rolling the block, rather than sliding it. By attaching 12 identical wooden rods to the faces of the block, one effectively transforms the block into a dodecagon prism with very little added mass, much lower ground pressure, and with good cross country mobility. The results are consistent with other researchers who attached wooden quarter circle "cradle runners" to full sized blocks, effectively turning them into cylinders. The effective dynamic coefficient of friction is found to be approximately of $\mu_d = 0.30 \pm 0.05$ for steady rolling motion. The force needed to maintain this motion is only $(\mu_d\, mg)/2$, because of the 2:1 mechanical advantage inherent in the use of the rope being wrapped around the block. So, to move the block requires that the crew be able to apply a force of only $0.15*mg$. The crew does, however, need to walk twice as far as the block is moved. It would seem that some variation of rolling the blocks should now be considered to be among the "best" and most likely method used to move the stones for the great pyramids.

## V. ACKNOWLEDGEMENTS

The authors would like to thank the following for assistance: Erica O. West, Wilson Middle School, Terre Haute, IN for initial discussions and experiments with rolling blocks, Joseph Pettit, ISU Department of Geology for advice on the lashing of rods, Scott Tillman, University Architect for materials, Lori Vancsa, Office of Environmental Safety for safety equipment, and Stephanie Krull, Facilities Management Department for access to the experimental locations. In addition, the authors would like to thank the following for financial support, ISU College of Arts and Sciences (GG), ISU Department Chemistry and Physics (KW), ISU Summer Undergraduate Research Experience (SURE 2012).

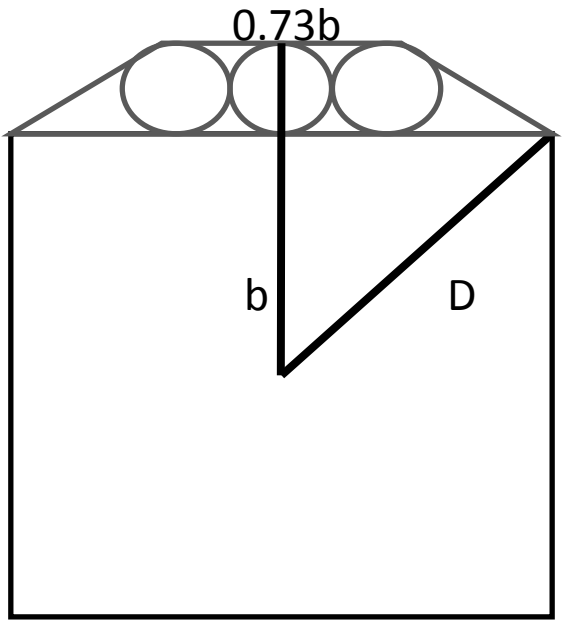

Figure 1. By adding three identical rods to one face of the square prism, that face is transformed into 3 identical faces of the dodecagon prism. The ideal rod diameter is D – b = 0.366 b, where b is from the center of the square to the face of the square, or half the length of one side of the square. Each face is 0.73 b in length.

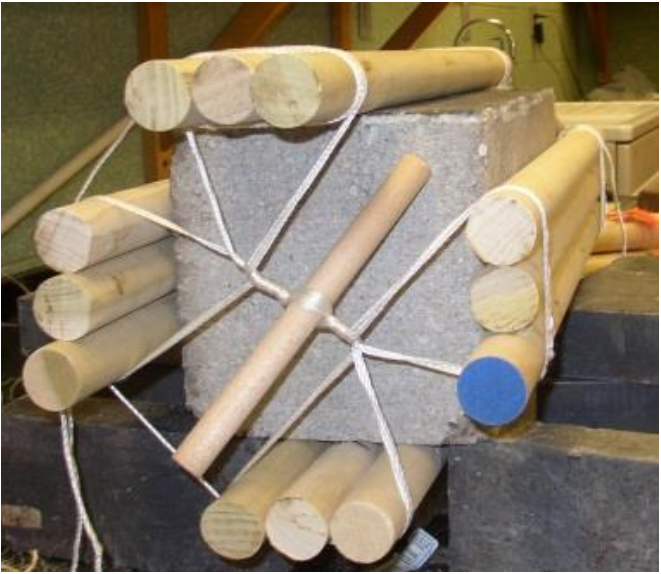

Fig. 2. The concrete block with the 12 rods attached. A second variation, with smaller diameter rods placed in the two intersection of the three larger rods also proved stable, and was successful in reducing the peak force required to start the rolling motion of the block.

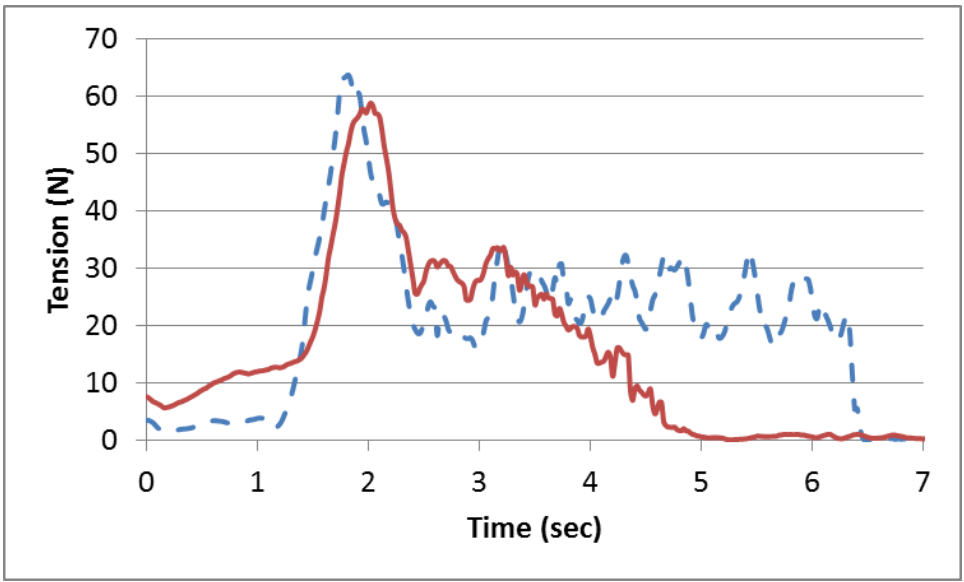

3a

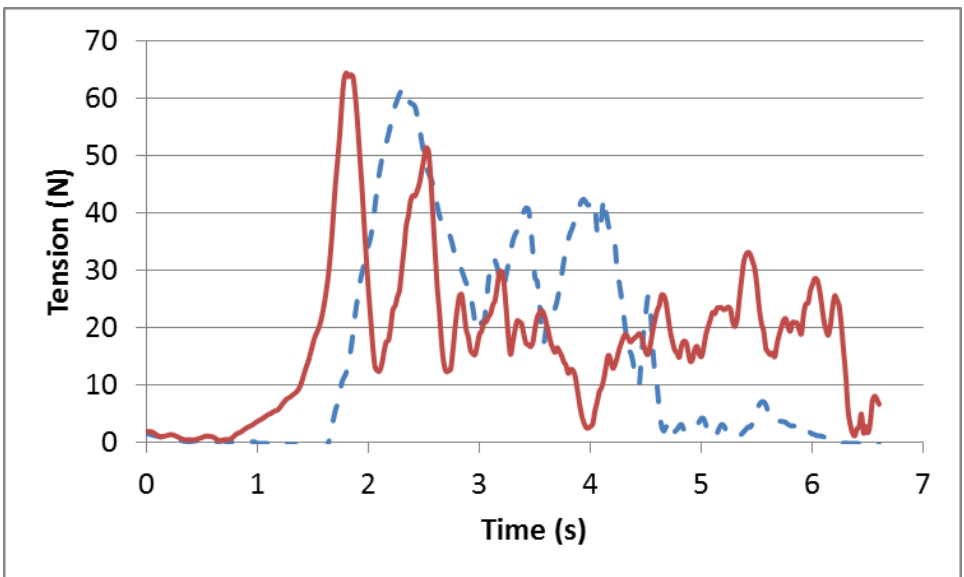

3b

Fig. 3. The data obtained by rolling the concrete block in Grass (3a dashed line) on the Infield (3a solid line), in Sand (3b dashed line) and on gravel (3b solid line). With mg = 290 N, a Tension of 50 N requires work equivalent to $\mu_d = 0.345$. The weight of the block, not including rods or rope, is 290 N.